\documentclass[lettersize,journal]{IEEEtran}
\usepackage{amsmath,amsfonts}
\usepackage{algorithmic}
\usepackage{algorithm}
\usepackage{array}
\usepackage{amssymb}
\usepackage[caption=false,font=normalsize,labelfont=sf,textfont=sf]{subfig}
\usepackage{textcomp}
\usepackage{stfloats}
\usepackage{url}
\usepackage{verbatim}
\usepackage{graphicx}
\usepackage{cite}
\usepackage{newunicodechar}
\usepackage{arydshln}
\begin{document}

\makeatletter
\newcommand{\linebreakand}{%
  \end{@IEEEauthorhalign}
  \hfill\mbox{}\par
  \mbox{}\hfill\begin{@IEEEauthorhalign}
}
\makeatother

\title{A Hypergraph-Based Machine Learning \\ Ensemble Network Intrusion Detection System}

\author{
    \IEEEauthorblockN{Zong-Zhi Lin\IEEEauthorrefmark{1}, Thomas D. Pike\IEEEauthorrefmark{1},
    Mark M. Bailey\IEEEauthorrefmark{1},
    Nathaniel D. Bastian\IEEEauthorrefmark{2}\IEEEauthorrefmark{1}
    }
    
    \IEEEauthorblockA{\IEEEauthorrefmark{1}Department of Cyber Intelligence \& Data Science, National Intelligence University, Bethesda, MD, USA
    \\ alexzzlin@gmail.com, \{thomas.d.pike, mark.m.bailey\}@odni.gov}
    
    \IEEEauthorblockA{\IEEEauthorrefmark{2}Army Cyber Institute, United States Military Academy, West Point, NY, USA
    \\ nathaniel.bastian@westpoint.edu}

    \thanks{This work has been submitted to the IEEE for possible publication. Copyright may be transferred without notice, after which this version may no longer be accessible.}
}
%



\maketitle

\begin{abstract}
Network intrusion detection systems (NIDS) to detect malicious attacks continue to meet challenges. NIDS are often developed offline while they face auto-generated port scan infiltration attempts, resulting in a significant time lag from adversarial adaption to NIDS response. To address these challenges, we use hypergraphs focused on internet protocol addresses and destination ports to capture evolving patterns of port scan attacks. The derived set of hypergraph-based metrics are then used to train an ensemble machine learning (ML) based NIDS that allows for real-time adaption in monitoring and detecting port scanning activities, other types of attacks, and adversarial intrusions at high accuracy, precision and recall performances. This ML adapting NIDS was developed through the combination of (1) intrusion examples, (2) NIDS update rules, (3) attack threshold choices to trigger NIDS retraining requests, and (4) a production environment with no prior knowledge of the nature of network traffic. 40 scenarios were auto-generated to evaluate the ML ensemble NIDS comprising three tree-based models. The resulting ML Ensemble NIDS was extended and evaluated with the CIC-IDS2017 dataset. Results show that under the model settings of an Update-ALL-NIDS rule (specifically retrain and update all the three models upon the same NIDS retraining request) the proposed ML ensemble NIDS evolved intelligently and produced the best results with nearly $100\%$ detection performance throughout the simulation.
\end{abstract}


\section{Introduction} \label{introduction}
\IEEEPARstart{T}o defend against unauthorized access an effective cybersecurity system must be supported by machine generated responses to keep pace with a diverse and expanding array of attackers, who are supported by suites of AI enabled tools. The increase in network size and network activities pose considerable challenges for the limited human resources and their comparatively slow pace to adapt against cyberthreats. Furthermore, the rapid advances in internet and communication technologies continue to introduce increasingly complex dynamics comprising of more variants of cyberthreats, such as malware, phishing, port scanning, and denial-of-service attack, to name just a few. These rapidly evolving network security threats and the need to detect, characterize and defend against diverse and adapting cyberattacks necessitate the integration of machine learning techniques into cybersecurity practices \cite{DeLucia_Maxwell_Bastian_eta_2021}. 

Machine learning (ML) techniques can be categorized into the two seminal categories as either supervised or unsupervised learning algorithms \cite{Devi_Abualkibash_2019}. Supervised learning algorithms learn from labeled data (e.g. data labeled as known intrusion attempt), whereas unsupervised learning algorithms statistically assess unlabeled data, based on the the features of that data. These ML methods often require large amounts of network traffic data with laboriously developed data features to train network intrusion detection systems (NIDS)\cite{Maxwell_Alhajjar_Bastian_2019,Chale_Bastian_2021}. To overcome this training data challenge, transfer learning can be employed to leverage the curated data from network traffic and apply it to another, without necessarily needing to use the same ML approach \cite{Bierbrauer_DeLuciab_Reddya_2022}.

ML-based NIDS continue to struggle to develop approaches that effectively integrate ML into NIDS \cite{Bierbrauer_Kritzer_Chang_Bastian_2021,Devine_Bastian_2021}. A grave challenge to NIDS has been its vulnerabilities to ML generated attacks that escape the detection with a high level of success \cite{Alhajjar_Maxwell_Bastian_2021,Schneider_Aspinall_Bastian2021}.  Furthermore, NIDS involving ML are often trained and developed offline causing a time lag in fixing vulnerabilities as attacks spread across the network \cite{Bielawski_Gaynier_etal_2020}. ML based approaches have proven effective at developing increasingly successful attack vectors, but struggle to develop effective defense approaches. The increasingly heterogeneous and complex dynamics of the cyber-attacks, many produced through ML approaches, requires cyber-defenses to effectively integrate ML into their strategies.  

 A fundamental challenge for developing an effective ML based NIDS is overfitting. In order for a deployed ML-based NIDS model to generalize to novel attacks, it must be optimally tuned to prevent it from performing too well on its training data. \cite{Aljanabi_Ismail_Ali_2021}. The reality in ML is that one should never trust a ML model that gets 100\% of its training data correct because that model will perform very poorly when faced with novel attacks, a phenomenon known as the bias-variance trade-off. This poses a significant risk to network security, and presents a daunting challenge in devising effective ML-based NIDS to defend against ML based attacks.

Our work seeks to aid NIDS development through the integration of a hypergraph metrics and a ML ensemble approach focused on port scan attacks. Integrated into this approach is auto generated novel attacks the NIDS must continue to identify and defend against. First, we use a variety $s$-closeness centrality metrics to capture the adversary's signature pattern of clustered port scan activities on a targeted machine. Second, we investigate multiple ways to render these hypergraph metrics as features augmenting the NIDS training dataset. Finally, we devise a methodology to create an intelligent cybersystem to continuously train and update a ML ensemble NIDS in real-time within the modeled network.

\section{Related Works} \label{lit_review}

ML models are applied across a variety of domains to include  cybersecurity, with varying effectiveness. In the context of digital world, the attacker launches an attack on victims through the computer network; therefore, it is natural to research cybersecurity from the perspective of a network problem \cite{Xiao_Liu_etal_2020}. One can model the network traffic data from source and destination internet protocols (IPs), source and destination ports as nodes in a network, while the protocols and attacks can be modeled as links. Network science based approaches have proven effective in understanding the system traffic to find attack information, detect abnormal traffic, or devise network detection and prevention approaches \cite{Khraisat_Gondal_etal_2019}.

Although network science is ubiquitous in cybersecurity, limited work explores the utility of hypergraphs, which is an extension of network science to algebraic topology, to aid cybersecuirty efforts and in this case, specifically the intrusion detection problem. Earlier studies model the multi-step of logged cyber activities (e.g., remote access, directory scanning, SQL injection and web Server Control to attain denial of service goal) as a hypergraph to enable the detection of sub-sequences that represent an intrusion \cite{Wang_Liu_Jajodia_2006,Guzzo_Pugliese_etal_2014}. The authors in \cite{Guzzo_Pugliese_etal_2014} showed that the intrusion detection problem defined over their cyberattack model is a NP-Complete problem.

A few network intrusion studies have used hypergraphs to improve the performance of ML-based NIDS to detect several specific types of intrusions such as port scan, Brute-force, Heartbleed, Botnet, Denial of Service, Web attacks, and Distributed Denial of Service (DDoS). The study in \cite{Raman_Kannan_etal_2016} uses hypergraphs to identify the key features for NIDS by way of modeling the multi-way relationships among minimal sets of explanatory attributes resulting from dimensionality reduction with respect to the degree of dependency among the explanatory attributes. The study in \cite{Raman_Somu_2017} exploits properties of hypergraphs to optimize its genetic algorithm performance for parameter setting and feature selection in a support vector machine model to balance evaluation criteria of intrusion detection rate, false alarm rate and number of features. The hypergraph-based study in \cite{Xiao_Liu_etal_2020} uses a hypergraph as an intermediate construct to derive a number of latent features, but not to reveal concerning patterns of network activities. The study in \cite{An_Su_Lu_Lin_2018} proposes a hypergraph clustering model based on the \textit{a priori} algorithm to describe the association between fog nodes which are suffering from the threat of DDoS, thereby increasing resource utilization rate of the system. The authors in \cite{Joslyn_Aksoy_Arendt_etal_2020} uses a hypergraph approach using Domain Name Systems cyber data to represent the joint relationships between collections of domains (coded as hyperedges) and IP addresses (coded as vertices), enabling both analytic and visual explorations of its $s$-components and distributions of these component sizes in terms of node and edge counts to find abnormal IPs and domains and examine the neighborhood of known bad IPs or domains. However, \cite{Joslyn_Aksoy_Arendt_etal_2020} did not use hypergraph-based metrics as part of the features to train ML NIDS yet. 

Another facet of NIDS challenges is the need for automated updates and deployment to rapidly mitigate vulnerabilities and stop the spread of infiltrated attacks into other parts of the network. ML based approaches are uniquely suited to support automatic updates The study in \cite{Kalekar_Kshatriya_etal_2014} implements a Naïve Bayesian classifier to balance detection for different types of real-time networking attacks such as majority intrusion of denial of service and minority intrusion of user to root in real network connections datasets. The study in \cite{Sangkatsanee_Wattanapongsakorn_2011} used a decision tree and selected 12 essential features of network to implement a real-time intrusion detection system to distinguish normal network activities from main attack types (Probe and Denial of Service) with a detection rate higher than 98\% within 2 seconds. The study in \cite{Thirimanne_Jayawardana_2022} implements a real-time NIDS based on a deep neural network to identify intrusions by analyzing network data in real-time, achieving at 81\% for accuracy in testing performance results. The challenge is that the real-time ML NIDS in these studies and many others continue to miss some intrusions at some small likelihood.

As ML becomes integrated into cyberattacks an additional aspect any NIDS must consider is the use adversarial training to automatically adapt intrusion attempts to evade detection and be classified as legitimate or normal\cite{Shipp_Clouse_eta_Bastian_2020,Bierbrauer_eta_Bastian_2021}. Optimization models have been rigorously applied in this data augmentation approach but have shown limited success in enhancing cyberdefense \cite{Kolter_Madry_2019}. In addition, experiments have also shown evolutionary approaches in constrained environments to generate adversarial examples to evade NIDS, mimicking the learning behavior of potential attackers to effectively cause high misclassification rates for more than ten commonly used ML models \cite{Alhajjar_Maxwell_Bastian_2021}.

Previous research has used hypergraphs and ML in various forms and with varying degrees of sophistication to enhance cybersecurity. From a more general perspective, these research efforts have a common thread that treats robustness (the ability to withstand an attack) and resilience (the ability to recover after being successfully attacked) as mutually exclusive goals, while generally favoring robustness. Our research combines hypergraph metrics and ML automation, in an effort to develop and evolve a NIDS that can both withstand attacks, but also recover quickly when penetrated.

\section{Evaluation Methodology} \label{research_method}

To devise and evaluate the novel ML Ensemble NIDS, our evaluation methodology comprises several connected components, including (A) NIDS Evaluation Framework consisting of updating NIDS Detection Performance with Hypergraph and updating NIDS Adversarial attacks, (B) NIDS Model Performance Evaluation, and (C) Research Dataset. These components allow for the iterative simulation of the evolutionary fight of the deployed NIDS models and the adapting cyber-attacks using generative adversarial examples. 

\subsection{NIDS Evaluation Framework}

A set of virtual computer agents connected with simple rules of interaction simulates the computer network used for this research, as illustrated with the network of three computers in the green circle in Figure 1. The Ensemble NIDS Model within the network takes and renders the features of network traffic data to determine if it is an intrusion. To simplify the interactions among computers, the computers deploy with the same set of NIDS, accordingly each can be updated with the same retrained NIDS or the set of retrained NIDS.

The Adversarial Examples Generation module applies state-of-the art techniques to generate and introduce adversarial examples to simulate the unseen adversarial attacks to the network of computers. This enables data augmentation for the NIDS model training, thereby auditing or enhancing the adversarial robustness of deployed NIDS models. Successful intrusions are sent to the Network Traffic Database module to serve as part of train dataset to retrain NIDS models.

The Network Traffic Database module collects network traffic data, NIDS classified ``Attack” samples, and failed-to-detect adversarial examples. Whenever a request to re-train any NIDS model is triggered, the Network Traffic Database module provides a new training dataset mixing normal network traffic data adequately proportionate to the size of cumulative evaded successful intrusions at the time of NIDS model retraining request.

The Behavioral Analytics module applies hypergraph analysis to provide systemic oversight over a longer time span at each computer or the entire network to detect intrusions sometime after the port scan infiltration.

The NIDS Model module re-trains and updates NIDS to improve detection performance level, whereas the Computer Module maintains an ensemble of ML-based NIDS to differentiate normal network users from cyber-attacks. Note that infiltrated cyber-attacks in a computer can spread to other computers within the simulated network. The Network of Computers module coordinates the NIDS updates from the NIDS retraining request at other computers within the network \cite{Stonedahl_Wilensky_2008} and coordinates with the Behavioral Analytics module to keep the ML Ensemble NIDS updated.

The Scorecard module collects and compute the performance of deployed NIDS in terms of the total of infiltrated port scans and failed-to-detect adversarial examples and its associated confusion matrix at each computer over time. When the total of infiltrated port scans exceed a specified threshold choice, a request to re-train and update a NIDS model or a NIDS ensemble model will be triggered to enable evidence-based decisions and maintain and/or improve performance of deployed NIDS models in the intrusion detection accuracy. The deployed ML Ensemble NIDS models continue to defend the system if there is no retraining request. The evaluation process continues and stops when it reaches the specified simulation time.

\begin{figure*}[ht]
\vskip 0.2in
\begin{center}
\centerline{\includegraphics[width=1.85\columnwidth]{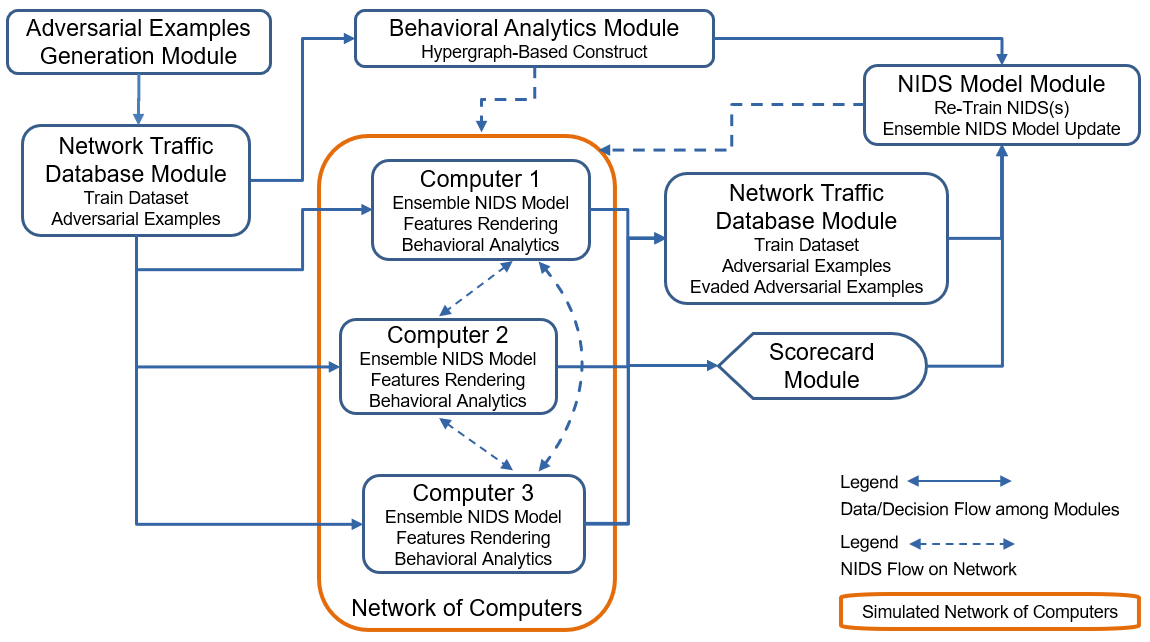}}
\caption{Evaluation Framework for the ML Ensemble NIDS}
\end{center}
\label{ieee-ensemble-nids}
\vskip -0.2in
\end{figure*}

\subsection{NIDS Model Performance Evaluation}

ML models are typically developed in two stages; a training stage to develop the model and a testing stage to evaluate the developed model. Likewise, the research data is split using the 80\% and 20\% convention between training and testing stages. Individual outputs of ML models are generally classified using numerical likelihood. If the classification is a binary choice of A and B, then when the likelihood is very close to 1.0 for category A, the interpretation is that the model predicts Category A to realize with very high confidence. If the likelihood is very close to 0.0, the model predicts Category A to happen with very low confidence. If the likelihood falls around 0.5, implying high uncertainty, this often requires further review of the data to support the eventual prediction. Built on raw likelihood scores, we adopt the commonly used confusion matrix and mean $F_1$-score to evaluate the performance of the trained ML models as well as the proposed ML Ensemble NIDS.

\subsection{Research Dataset}

We use the intrusion detection evaluation dataset, publicly released in 2017 from the Canadian Institute for Cybersecurity (CIC-IDS2017). The CIC-IDS2017 data set consists of several attack classes, to include: port scan, Brute-force, Heartbleed, Botnet, Denial of Service, Distributed Denial of Service, Web attacks, and infiltration of the network from inside \cite{cic_ids_2017,Sharafaldin_Lashkari_etal_2018}.

Our work considers the port scan attack class. Out of its total $84$ features in the CIC-IDS2017 port scan dataset, eight (8) of them are understood to be most important features on detecting attacks: (1) duration of the flow in microsecond, (2) total packets in the forward direction, (3) total packets in the backward direction, (4) total size of packet in forward direction, (5) total size of packet in backward direction, (6) number of flow bytes per second, (7) number of flow packets per second, and (8) download and upload ratio \cite{Xiao_Liu_etal_2020}. Out of its total 286,487 labeled records, total 407 records with negative duration of the flow in microsecond, or missing or infinite values for at least one of those eight features were removed from the research dataset. The ratio of labeled port scan attacks and benign records in the resulting research dataset of 286,060 labeled records is about $55.5\%$ and $44.5\%$, respectively.

The feature of transport protocol is included as a categorical variable because the dataset indicates the number of port scan events via the transport protocol six (namely, Transmission Control Protocol, TCP) is $158,797$, which is about 2.37 times of the 67,053 benign events via the same transport protocol. Two other values of the protocol feature are 0 (version 6 of the Internet Protocol, IPv6) and 17 (User Datagram Protocol, UDP). We use the set of nine (9) network raw features (NRF, the protocol feature and the eight features listed above) as the base training dataset to train the base NIDS model in the ML Ensemble NIDS. As depicted by Table \ref{research_dataset_numbers} summarizing basic descriptive statistics for each numeric feature of this base training dataset NRF by record type of Port Scan and BENIGN. There exists noticeable differences in the statistics among those eight features to provide their differentiating power to detect network activity type. This set NRF will be expanded further with hypergraph-based features to be derived later based on another three network features of source ip, destination ip, and destination port number. Source port number is typically randomly assigned, thus it is not included for further uses.

\begin{table*}[ht]
\caption{Selected Features Descriptions from CIC-IDS2017 port scan data file to define the Research Data Set.}
\label{research_dataset_numbers}
\begin{center}
\begin{small}
\begin{sc}
\begin{tabular}{lcrrrr}
\hline
Feature Description & Data Type & mode & maximum & average & std. dev. \\
\hline
duration of the flow in microsecond & Port Scan & 47 & 119809735 & 82885.94 & 2326775.89 \\
 & Benign & 30985 & 119999949 & 5386984.20 & 31562986.85 \\
\cline{2-6}
Total packets in the forward direction & Port Scan & 1 & 150 & 1.02  & 0.43 \\
 & Benign & 2 & 3119 & 6.55  & 28.98 \\
 \cline{2-6}
Total packets in the backward direction & Port Scan & 1 & 30 & 1.00  & 0.15 \\
 & Benign & 2 & 3635 & 6.67  & 42.23 \\
 \cline{2-6}
Total size of packet in forward direction & Port Scan & 0 & 1473 &  1.09 & 5.66 \\
 & Benign & 68 & 232349 &  524.05 & 2771.69 \\
 \cline{2-6}
Total size of packet in backward direction & Port Scan & 6 &  11595 & 12.24  & 267.40  \\ 
 & Benign & 142 &  7150819 & 6079.02  & 76351.31  \\ 
 \cline{2-6}
number of flow bytes per second & Port Scan & 139535 & 8000000 & 220359.75  &  459863.69	\\
 & Benign & 5684 & 2070000000 & 2241033.31  &  38472283.25	\\
 \cline{2-6}
number of flow packets per second & Port Scan & 42553 & 2000000 & 62690.57  & 127930.00 \\
 & Benign & 113 & 3000000 & 62501.33  & 247879.03 \\
\cline{2-6}
download and upload ratio & Port Scan & 1 &  2 & 0.99 & 0.09 \\
 & Benign & 1 &  124 & 0.67 & 0.62 \\
\hline
\hline
\end{tabular}
\end{sc}
\end{small}
\end{center}
\vskip -0.1in
\end{table*}

Two additional practices were applied to expand, diversify, and toughen the port scan attack dataset to improve the development of the proposed ML Ensemble NIDS.
\begin{itemize}
 \item Leverage the CIC-IDS2017 dataset to generate a dataset simulating the case where multiple hacker source IPs port scans multiple destination IPs.
 \item Apply an adversarial example generation technique on the CIC-IDS2017 dataset to enable data augmentation needed to conduct adversarial training.
\end{itemize}

\section{Developing a Hypergraph-Based \\ ML Ensemble NIDS}

Network traffic flow data typically includes source and destination IPs, source and destination ports, and the communication protocol, as found in Network Intrusion Detection Evaluation Dataset CIC-IDS2017 \cite{cic_ids_2017} used for this study. This data was then abstracted into a hypergraph to capture the pattern of port scan activities. Mirroring the concept of using closeness centrality to measure the proximity of one node to all others in a network, the abstracted hypergraph enables to formulate a set of the $s$-closeness centrality to quantify the pattern. These results create a set of hypergraph metrics to use in training the NIDS models, thereby becoming the building blocks of the proposed ML Ensemble NIDS. We then extend these results to develop a behavioral analytic capability monitoring the network activities with no prior knowledge of their activity types to assist with online detection of port scan activities. It is worth noting that modeling the multi-dimensional cyber-security data as hypergraphs, the visualization of hypergraphs and the computation of hypergraph metric features were accomplished using the HyperNetX library \cite{Monson_Arendt_eta_2019, Praggastis_Arendt_eta_2022}.

\subsection{Model Port-Scanning Activities with Hypergraph}

A basic understanding of hypergraphs is necessary to appreciate the metrics of interest. A hypergraph comprises a set of vertices $\mathcal{V}$ and set of hyperedges $\mathcal{E} = \{ e_1,e_2,\dots,e_m \}$ such that each hyperedge is a subset of vertices set $\mathcal{V}$, namely $e_i \subseteq \mathcal{V}$. A typical graph or network is a hypergraph where all its hyperedges have size of two vertices, i.e., $|e_i|=2$. An $s$-walk of length $k$ between hyperedge $f$ and $g$ is a sequence of hyperedges $f=e_{i_0},e_{i_1},e_{i_2},\dots,e_{i_k}=g$ such that for $j=1,\dots,k$, we have $\vert e_{i_{j-1}} \cap e_{i_j} \vert \geq s$ and $i_{j-1} \neq i_j$ \cite{Aksoy_Joslyn_eta_2020}. An $s$-path is an $s$-walk where hyperedges are not repeated. The $s$-distance between hyperedge $e$ and $f$ is the length of its shortest $s$-path, denoted as $d_s(e,f)$. In addition to length, each $s$-path in a hypergraph has width, the minimal number of common vertices among all the consecutive hyperedges of the $s$-path. In other words, the width of an $s$-path is at least $s$ since all consecutive intersections are at least size $s$. An $s$-component in a hypergraph is a collection of hyperedges so that any pair of hyperedges are connected via an $s$-path.

In the context of port scanning activities, let $S_{s-ip}$, $S_{d-ip}$ and $S_{d-port}$ be the sets of all source IP addresses, all destination IP addresses and all destination ports, respectively. Let $\mathcal{E}$ be a set of hyperedges in which each hyperedge is either source or destination IP address that is associated with port scanning a set of destination ports in a stream of network activities. The constructed hypergraph can be denoted as $HG = \{ \mathcal{V} = S_{d-port}, \mathcal{E} = S_{s-ip} \cup S_{d-ip} \}$, so each unique IP or port is only included once in the hypergraph. Thus, this constructed hypergraph summarizes the structure of interactions and connectivity among IPs and destination ports involved in network activities modeled.

Each hyperedge in $\mathcal{E} = S_{s-ip} \cup S_{d-ip}$ is mathematically defined as a subset of $S_{d-port}$ based on its associated destination port scanning activities in a network. For example, the edge of source IP-173.194.208.155 at the top right corner of Fig. 2 
comprises a single node of destination port 35066 reflecting its access in a normal network activity. As depicted in the Fig. 2 with a small hypergraph example, the pair of Source IP-172.16.10.1 and Destination IP-192.16.10.50 addresses were associated  with a long and common list of available destination ports, the crowded black dots in Fig. 2, 
involved in the network activities modeled \cite{Sharafaldin_Lashkari_etal_2018}.
It happens that the attacker from source IP 172.16.0.1 port scanned the targeted destination IP 192.168.10.50 via a list of destination ports, including ports 53 and 80 that were used normally by other source IPs.

\begin{figure*}[ht]
\vskip 0.2in
\begin{center}
 \centerline{\includegraphics[width=1.9\columnwidth]{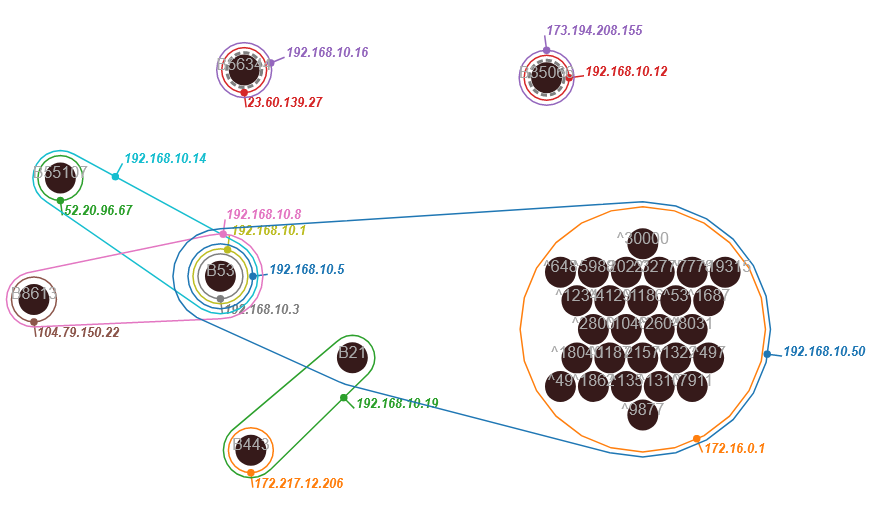}}
 \label{hypergraph-examples-43}
 \caption{This hypergraph with 15 edges and 34 vertices was constructed by 43 records with benign and port scan attack class label denoted as B and $\wedge$, respectively. The ports 21, 53, 443, 56344, 8613, 35066, 42154, and 55107 in normal uses by source and destination IPs were also part of smaller set of centers to concentric circles associated with source or destination IPs.}
\end{center}
\vskip -0.2in
\end{figure*}

The $s$-closeness centrality $C_s(e)$ for an edge $e$ measures how close $e$ is to all others in its $s$-component. It is computed by the following formula $C_s(e) =\frac{|\mathcal{E}_s|-1}{\sum_{f \in \mathcal{E}_s} d_s(e,f)}$, where $\mathcal{E}_s$ is the set of hyperedges in its $s$-component that contains the edge $e$ \cite{Aksoy_Joslyn_eta_2020}. It can be readily verified that the $s$-closeness centrality for each hyperedge in the $s$-component comprising only concentric circles is one. As seen in Fig. 3,
for $s \in \{ 11, 12,\dots,30 \}$, $C_s(e \in \{ 172.16.0.1, 192.168.10.50 \}) \rightarrow 1$, and the two edges of IP addresses almost form two concentric circles based on the set of commonly accessed destination ports. Coincidentally, the source IP is the hacker used to scan the destination ports available to the destination IP in the training dataset. Therefore, these individual $s$-closeness centrality metrics essentially capture the hacker’s signature pattern of port scan activities on a targeted machine and have potential to render features to expand the NIDS training dataset.

We remark the $s$-closeness centralities for the benign network activities vary between zero and one before the value of $s$ hits $22$. After it hits $22$, the majority of them stay at zero and some at one, thereby yielding a decreasing trend of its mean $s$-closeness centralities to close to zero, as indicated by the blue curve in Fig. 4. To expand the proposed ML Ensemble NIDS to detect both benign and other types of attacks with high performance, this decreasing trend of its mean $s$-closeness centralities proves in our numerical evaluations to be effective in encoding the pairs of non-hacker IPs with a set of weights approximating the distribution of their average s-closeness centralities. 


\begin{figure}[ht]
\vskip 0.2in
\begin{center}
 \centerline{\includegraphics[width=0.9\columnwidth]{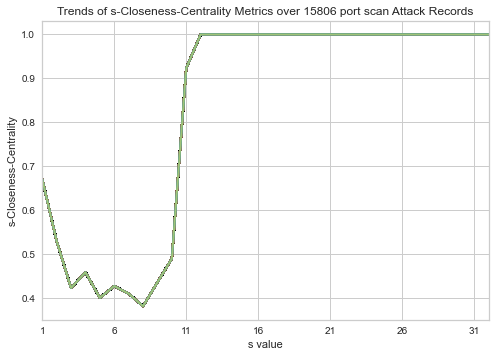}}
 \label{hypergraph-measures}
 \caption{Trends of $s$-Closeness-Centrality ($s$-C-C) metrics of hyperedges for port scan records in a hypergraph whose hyperedges are defined as source or destination IP address associated with the nodes of scanning destination ports. This trend of $s$-Closeness-Centrality metrics chart is for hyperedges of destination IP based on $15,806$ port scan only records of CIC-IDS2017 port scan dataset.}
\end{center}
\vskip -0.2in
\end{figure}

\begin{figure}[ht]
\vskip 0.2in
\begin{center}
 \centerline{\includegraphics[width=0.86\columnwidth]{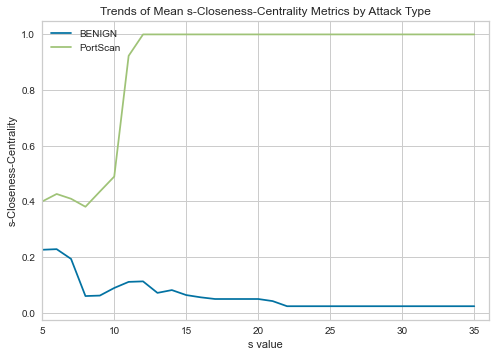}}
 \label{hypergraph-measures-average}
 \caption{Trends of mean $s$-Closeness-Centrality ($s$-C-C) metrics of hyperedges by Attack Type in a hypergraph whose hyperedges are defined as source or destination IP address associated with scanning the nodes of destination ports. These trends of mean $s$-Closeness-Centrality metrics chart by Attack Type are for hyperedges of destination IP based on $28,645$ records (Benign or port scan) of CIC-IDS2017 port scan dataset.}
\end{center}
\vskip -0.2in
\end{figure}

\subsection{Generating Hypergraph-Based Feature Sets}

We now formulate the mathematical strategy to compute $s$-closeness centralities of the edges in the hypergraph with a properly specified skip-interval $k$. The strategy creates two different sets of hypergraph-based features to use in training the NIDS models. Both feature sets augment the dataset of the aforementioned nine raw network flow features.

Let $C_{3+n \times k}(e)$ be the $(3+n \times k)$-closeness centrality of the hyperedge $e$ for a positive integer $n$. It often takes much longer time to compute $C_1(e)$ and $C_2(e)$ (especially for a much larger hypergraph), yet when $s \in \{ 1,2 \}$, the first two $s$-closeness centralities are more likely to contain signals from the normal uses of the destination ports by other hyperedges, thereby statistically causing the first two $s$-closeness centralities to reduce its differentiating power if there exists a pattern of port scan activities. In the numerical evaluation, let the values of $n \in \mathcal{S} = \{ 0, 1, 2, \dots, 9, 10 \}$ and the value of skip-interval $k$ is subject to the condition that the value of ($3+10 \times k$) is about $70\%$ of the maximal total number of nodes or destination ports connected to hyperedges (namely, source IP or destination IP).

It is by design to evenly spread out the computation of these $s$-closeness centralities because the values of latter $s$-closeness centralities (namely at larger $s$) carry more signal if there exists a pattern of port scan activities. For computational design convenience and completeness, let all the remaining $s$-closeness centralities be zero (0) at hyperedge $e$ when the value of $s$ is greater than its total number of nodes or destination ports but is smaller than the value of $3+10 \times k$.

Adopting the above mathematical definitions and symbols, we construct two different Hypergraph ($HG$) feature sets for training NIDS and their numerical evaluations:
\begin{itemize}
 \item Feature set $HGI$ includes the set of nine raw features and all the individually ($I$) computed $s$-closeness-centralities ($C_{3+n \times k}(e), n \in \mathcal{S}$) and their sum ($\sum_{n \in \mathcal{S}} C_{3+n \times k}(e)$) for each hyperedge $e$. The resulting data set $HGI$ is used to train the NIDS model $HGI$.
 \item Feature set $HGA$ includes the set of nine raw features, the last computed $n^{th}$ $s$-closeness-centrality (namely $C_{3+10 \times k}(e)$) and four (4) aggregate ($A$) hypergraph measures. The resulting dataset $HGA$ is used to train the NIDS model $HGA$. The four aggregate hypergraph measures are (1) the sum of all the computed $s$-closeness-centralities (i.e., $\sum_{n \in \mathcal{S}} C_{3+n \times k}(e)$), (2) the total number of unique destination ports associated with source IPs hyperedge (i.e., $\vert e_{s-ip} \vert$), (3) the number of destination ports associated with destination IPs hyperedge (i.e., $\vert e_{d-ip} \vert$), and (4) the sum of destination ports tied to source or destination IPs (i.e., $\vert e_{s-ip} \vert + \vert e_{d-ip} \vert$).
\end{itemize}

\subsection{Hypergraph-Based ML Ensemble NIDS Model}

The proposed hypergraph-based ML Ensemble NIDS model comprises three classifiers based on Random Forest and Light Gradient Boosting Machine techniques \cite{pycaret_rf_lightgbm_2020}:
\begin{itemize}
 \item The NIDS model $RF$ trained with the nine raw features in the train dataset using a Random Forest method.
 \item The NIDS model $HGI$ trained with the feature set $HGI$ using a Light Gradient Boosting Machine.
 \item The NIDS model $HGA$ trained with the feature set $HGA$ using a Light Gradient Boosting Machine.
\end{itemize}

It is worth noting that the features comprising the $s$-closeness centralities of hyperedges used to define the hypergraph-based features in $HGI$ and $HGA$ are derived by keying the associated IPs (considered known, either source IPs or destination IPs) to the IPs in each records, or set as zeroes for any new or unseen IPs. 

\subsection{Behavioral Analytics with Hypergraph}

We now extend the above hypergraph-based results to develop a behavioral analytic capability assisting with real time detection of port scan activities.

In real-world settings, almost none of the NIDS models can claim to detect all the cyber-attacks at a 100\% confidence level to avoid over-fitting their model. In addition, there is generally no prior knowledge of malicious or benign nature of network traffic to each user or computer node. To mitigate this vulnerability in the context of port scanning activities, we use the hypergraph construct discussed above to monitor the simulated network traffic and conduct on-line analysis to detect if the signature pattern of port scan activities exists.

The detection mechanism of port scan activities exploits the signature pattern illustrated in Figure 2 and 3(a). The two IP addresses form two concentric circles containing a long and varying list of nodes of destination ports, and their $s$-closeness centralities $C_s(e) =\frac{|\mathcal{E}_s|-1}{\sum_{f \in \mathcal{E}_s} d_s(e,f)} \rightarrow 1$, for $s \in \{ 11, 12,\dots,30 \}$ and $e \in \{ 172.16.0.1, 192.168.10.50 \}$. To transform this observation to work with a varying size of common nodes between any pair of source and destination IPs, the process in Algorithm~\ref{alg:example} outlines the adaptations in its implementation.

\begin{algorithm}[tb]
   \caption{Process to Detect Potential Port Scan Activities}
   \label{alg:example}
\begin{algorithmic}
   \STATE {\bfseries Step 1:} Construct the hypergraph with network traffic data of their source and destination IPs and destination ports.
   \STATE {\bfseries Step 2:} Compute its eleven ($11$) evenly spaced out $s$-closeness centralities for each edge $e$, namely $C_{3+n \times k}(e), n \in \mathcal{S}$.
   \STATE {\bfseries Step 3:} Make each of last six $s$-closeness centralities as zero if it is less than $0.95$ or $1$ otherwise.
   \STATE {\bfseries Step 4:} Flag port scan activities if the sum of last six converted $s$-closeness centralities is greater than or equal to two and the pair of source and destination IPs has not been flagged yet.
\end{algorithmic}
\end{algorithm}

The tail $s$-closeness centralities are of interest to indicate there is a clustering pattern of port scan activities; therefore, one can start the computation of 11 evenly spaced $s$-closeness centralities at $s=3$. For example, if the maximum number of nodes (or destination ports) for a certain edge in the hypergraph is $110$, then the 11 evenly spaced $s$-closeness centralities to compute are $C'_s$s, $s \in \{ 3, 13, 23, 33, 43, 53, 63, 73, 83, 93, 103 \}$.

It is shown numerically that the $s$-closeness centralities for $s \in \{ 1, 2 \}$ can take a very long time to compute in a large hypergraph. Thus the size of network traffic data used to construct a hypergraph to implement this behavioral analytics capability needs to evaluate and set properly in order to receive timely alerts and/or initiate NIDS re-training requests to update the ML Ensemble NIDS. In addition, the choices of both parameters in Step 3 and 4 were based on limited evaluations numerically, requiring thorough investigations before putting them into any operational uses.

\section{Evaluating Hypergraph-Based ML Ensemble NIDS}

Conducting the evaluation of the proposed ML Ensemble NIDS according to the simulation process depicted in Fig. 1, the Evaluation Framework for the ML Ensemble NIDS, this section comprises evaluating the effectiveness of adversarial examples to cause target NIDS to misclassify, NIDS models, and the ML Ensemble NIDS, respectively. The ML Ensemble NIDS comprises three pre-trained and active NIDS (namely, $RF$, $HGI$ and $HGA$) and retraining another NIDS when the total cumulative number of infiltrated attacks at each computer or entire network during the simulation exceeds the threshold choice such as two (2) attacks.

In general, the retrained NIDS will replace the worst of the active NIDS only if its detection performance based on certain metrics, such as accuracy, is better than the worst of the three active NIDSs, or referring it as Forgo-The-Worst NIDS. Another option is to retrain and replace all the three active NIDSs correspondingly and respectively, or referring it as Update-ALL-NIDS by position and model.

Numerical evaluations are performed in two different settings. The first and reduced setting evaluates the proposed ML Ensemble NIDS framework with attacks against the ML Ensemble NIDS from the port scan only network traffic data along with the set of adversarial examples generated. Normal and benign network traffic activities in the port scan dataset are not used for attacking but will be used to produce new training dataset with a balanced mixing of normal network traffic data and cumulative evaded successful intrusions at the time of NIDS model retraining request. The second and complete setting evaluates the ML Ensemble NIDS with all the attacks of the CIC-IDS2017 dataset and the set of adversarial examples generated.

\subsection{Adversarial Example Generation and Effectiveness}

To evaluate the NIDS model robustness, research on robustness has recently advanced to assessing how susceptible models are to adversarial manipulation \cite{Schneider_Aspinall_Bastian2021}. To account for adversary adaption our model uses the Zeroth-Order-Optimization (ZOO) attack method against it to generate new attack examples \cite{Nicolae_Sinn_etal_2019}. Studies show that training ML models on adversarial examples provides better performance \cite{Baluja_Fischer_2017}. Thus, this work uses adversarial generated data to address the feature vulnerabilities and make incremental improvements on the deployed NIDS models. This helped us examine and enhance the ML Ensemble NIDS adversarial robustness \cite{Alhajjar_Maxwell_Bastian_2021}. To simplify the development and testing of the ML Ensemble NIDS, we use only the reduced dataset of the port scan network traffic in the evaluation.

Our model approach does assume the adversary does not know that the ML Ensemble NIDS uses hypergraph $s$-closeness centralities metrics, and so the adapting NIDS is trained with the nine raw features (RF) using a Light Gradient Boosting Machine \cite{pycaret_rf_lightgbm_2020}. The ZOO attack, which estimates the gradients of the target model using zeroth order stochastic coordinate descent along with dimension reduction, hierarchical attack and importance sampling techniques, perturbs these raw features to attack the substitute NIDS and generate adversarial examples \cite{Nicolae_Sinn_etal_2019}. Specifically, the process in Algorithm~\ref{alg:adv-examples} outlines the steps to generate adversarial examples from a set of about $114,900$ Port Scan activities sampled from the CIC-IDS2017 Port Scan dataset, assuming Hacker does not know the model uses hypergraph $s$-closeness centralities metrics. Out of total $17,188$ adversarial examples generated via Algorithm~\ref{alg:adv-examples}, $14,799$ of them with predictive Port Scan probabilities greater than or equal to $0.55$ were used in the numerical evaluations.

\begin{algorithm}[tb]
   \caption{Render Adversarial Examples from Port Scans}
   \label{alg:adv-examples}
\begin{algorithmic}
   \STATE {\bfseries Step 1:} Normalize each of the nine RFs of Port Scan activities and apply 85\% vs. 15\% to split the dataset into train and test datasets.
   \STATE {\bfseries Step 2:} Use only the nine RFs to construct and fit the substitute NIDS model with Light Gradient Boosting Machine approach.
   \STATE {\bfseries Step 3:} Generate adversarial samples with ART Zeroth Order Optimization (ZOO) attack from the test dataset.
   \STATE {\bfseries Step 4:} Perform inverse transformation operations to rescale the normalized dataset to their original scale.
   \STATE {\bfseries Step 5:} Use the substitute NIDS model to compute predictive Port Scan probability for each adversarial example.
   \STATE {\bfseries Step 6:} Keep the adversarial examples whose predictive Port Scan probabilities are greater than or equal to $0.55$ and label these adversarial examples as port scan activities.
\end{algorithmic}
\end{algorithm}

To assess the effectiveness of these adversarial examples, these $14,799$ examples were fed into four different NIDS models (substitute NIDS and the deployed NIDS models of $RF$, $HGI$ and $HGA$) to compute and construct their respective distributions of port scan detection scores. Their respective results yield detection performances in the lowest performance to highest performance of Model B, A, C, and D, as depicted and summarized in Fig. 5. 
The ranking of the four models is based on the adoption of port scan threshold score 0.5 (the dotted yellow horizontal line) to adjudicate if a network activity is a port scan.

\begin{figure}[ht]
\vskip 0.2in
\begin{center}
 \centerline{\includegraphics[width=0.9\columnwidth]{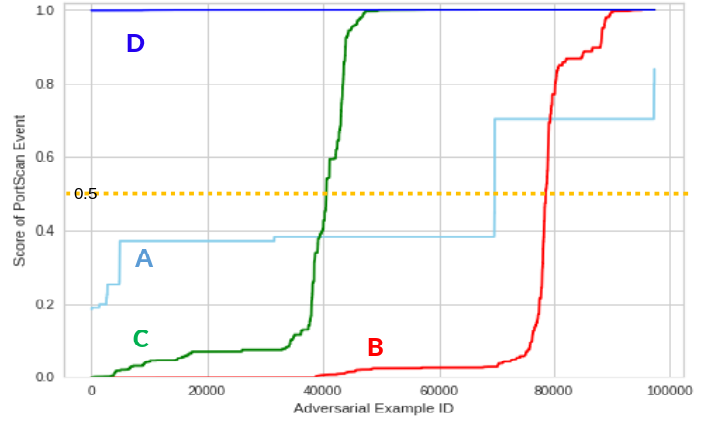}}
 \label{ieee-adv-examples-eval}
 \caption{Distributions of port scan Adversarial Examples Detection Score for four NIDS Models (A) Assuming Hacker does not know the model uses hypergraph $s$-closeness centralities metrics and uses only nine Raw Features (RFs) to construct its substitute NIDS model, (B) the NIDS model this work trained with nine RFs only, (C) the NIDS model trained with the dataset $HGI$, and (D) the NIDS model trained with the dataset $HGA$.}
\end{center}
\vskip -0.2in
\end{figure}

These curves of detection scores also suggest that there might have some combinations of those individually computed $s$-closeness centralities and their aggregates to include in the NIDS model training data set to reduce some portion of the left part in the chart for Model C. The curve in the chart for Model D essentially states that it does not need those ``considered and thoroughly evaluated” important raw features in the CIC-IDS2017 data set to help defend adversarial examples, given that the aggregate metrics of those $s$-closeness centralities metrics have captured the pair of IPs launching the port scanning attacks.

\subsection{Effectiveness of Hypergraph-Based NIDS Models}

To train a NIDS model for port scanning activities, as stated in an early section, nine out of 80 features in the CIC-IDS2017 port scan data file has been evaluated and commonly selected to be most important features on detecting attacks \cite{Xiao_Liu_etal_2020}. Therefore, we use the feature set $HGI$ comprising a set of hypergraph metrics and those nine features in the $40\%$ sample of the CIC-IDS2017 port scan dataset to train and build a tree-based NIDS. As indicated in Table ~\ref{model-perf-eval-table}, the derived set of hypergraph $s$-closeness-centrality metrics HGI enhances our NIDS model to reach nearly 100\% precision, recall and $F_1$ score performances.

\begin{table}[t]
\caption{Performances of tree-based classifiers on port scan over nine raw features, embedding features, and hypergraph $s$-metrics.}
\label{model-perf-eval-table}
\vskip 0.03in 
\begin{center}
\begin{small}
\begin{sc}
\begin{tabular}{lccc}
\hline
NIDS & Precision & Recall & $F_1$ \\
\hline
$RF$ Features*    & 0.9936 & 0.9912 & 0.9924 \\
$RF$+Embeddings* & 0.9998 & 0.9938 & 0.9968 \\
$RF$+$HGI$    & 1.0000 & 1.0000 & 1.0000 \\
\hline
\end{tabular}
\tiny{* Numbers in these two sections were obtained with the full CIC-IDS2017 Port Scan data [53].} \\
\end{sc}
\end{small}
\end{center}
\vskip -0.1in
\end{table}

\subsection{Evaluating the ML Ensemble NIDS with Port Scan data}

This research then evaluates the ML Ensemble NIDS in six cases with generic settings over a simulated network of 10 computers over 30 epochs. The threshold choice of infiltrated adversarial examples to start an NIDS re-training will be set and evaluated at each value in the set $TH = \{ 2, 5, 10, 20, 30, 40, 50, 100 \}$. Table ~\ref{Six_Scenarios} specifies the six scenarios in the following factors:

\begin{itemize}
 \item IP PAIRS: Pairs of source and destination IP.
 \item NIDS: Rule to update NIDS - Not Allowed, or Allowed to update and Forgo The Worst NIDS (FTW) or to Update All the NIDS models (UALL).
 \item ADV. Ex.: Include Adversarial Examples or not.
 \item THRESH.: Values of Threshold in the set $TH$ to trigger NIDS retraining request(s).
 \item PROD.: Emulate production environment with no prior knowledge of the nature of network traffic or not.
\end{itemize}

\begin{table}[t]
\caption{Designs of the six cases to evaluate the ML Ensemble NIDS}
\label{Six_Scenarios}
\begin{center}
\begin{small}
\begin{sc}
\begin{tabular}{cccccc}
\hline
Case & IP Pairs & NIDS & Adv. Ex. & Thresh. & Prod. \\
\hline
1 & 1  & Static & No  & $TH$  & No \\
2 & 16 & Static & No  & $TH$  & No \\
3 & 16 & FTW    & Yes & $TH$  & No \\
4 & 16 & UALL   & Yes & $TH$  & No \\
5 & 16 & UALL   & Yes & $TH$  & No \\
6 & 16 & UALL   & Yes & $TH$  & Yes \\
\hline
\end{tabular}
\end{sc}
\end{small}
\end{center}
\vskip -0.1in
\end{table}

Case 1 assumes knowing \textit{a priori} that port-scanning activities are from a known pair of source and destination IPs. The ML Ensemble NIDS stays static throughout the simulation. Numerical results indicate that the ML Ensemble NIDS yields the 0\% detection performance of False Negative Percentage (FNP) throughout the evaluation window, namely misclassifying zero Attack as Normal traffic. The NIDS $HGI$ and $HGA$ in the ML Ensemble NIDS help detect a small set of port scan activities that the NIDS $RF$ fails to detect, thereby enabling the ML Ensemble NIDS to attain perfect detection score and zero FNP throughout the simulation.


Case 2 extends Case 1 to include port scan activities from 15 other pairs of pseudo and unknown source and destination IPs. The ML Ensemble NIDS continues to stay static throughout the simulation.

The near 100\% detection performance of the ML Ensemble NIDS model in the FNP in Fig. 6
indicates that the NIDS $HGI$ and $HGA$ help detect most port scanning activities from multiple pairs of pseudo and unknown source and destination IPs that the NIDS $RF$ fails to detect as indicated by the Curve C in Fig. 5. 

It is seemingly the level of adversarial robustness in the ML Ensemble NIDS model is enhanced because the negative impact of simulated multiple pairs of unknown source and destination IPs on its detection performance is not as severe as originally expected from the observations in Fig. 6. This might be partly because the values of the nine (9) network flow features (NFFs) remain intact, and the tree-based NIDS based on these NFFs has a decent detection performance. A more realistic dataset of port scan attacks from unknown source to destination IPs is needed to study the impact of varying pairs of IPs on the detection performance of the ML Ensemble NIDS model.

\begin{figure}[ht]
\vskip 0.2in
\begin{center}
 \centerline{\includegraphics[width=0.9\columnwidth]{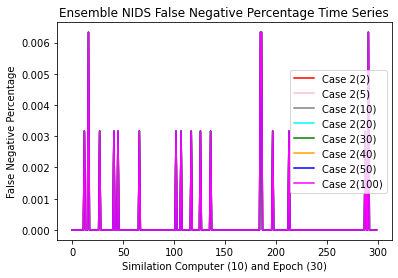}}
 \label{ieee-arnids-aide-s2}
 \caption{ML Ensemble NIDS Trend charts of FN ratio Series at each epoch and computer for Case 2.}
\end{center}
\vskip -0.2in
\end{figure}

Case 3 extends Case 2 to allow the retrained NIDS $HGI$ (when infiltrated events exceeds certain threshold choice) to replace at most one of the three active NIDS in the ML Ensemble NIDS with the Forgo-The-Worst (FTW) rule, namely when its detection rate of port scan events is higher than the worst of the three NIDS models. Its detection performance becomes somewhat volatile in the beginning and near the 150th run but stabilized after about 150th run, as depicted in Fig. 7. 

\begin{figure}[ht]
\vskip 0.2in
\begin{center}
 \centerline{\includegraphics[width=0.85\columnwidth]{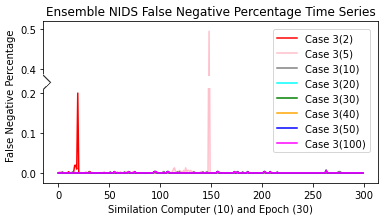}}
 \label{ieee-arnids-aide-s3}
 \caption{ML Ensemble NIDS Trend charts of FN ratio Series at each epoch and computer for Case 3.}
\end{center}
\vskip -0.2in
\end{figure}

Case 4 extends Case 3 to include attacks from adversarial examples generated with the ZOO attack method. Its detection performance greatly deteriorates in the beginning, as depicted in Fig. 8.
The deterioration causes the number of infiltrated attacks to exceed the study threshold choice in $TH$, thereby triggering the first NIDS HGI retraining request to replace the active NIDS with the lowest detection rate according to the FTW rule. After the first NIDS retraining and replacement, its detection performance gradually becomes somewhat stable and maintains a higher detection rate. In addition, the set of adversarial examples continues to infiltrate the ML Ensemble NIDS, thereby triggering more NIDS HGI retraining requests to replace and update the active NIDS in the ML Ensemble NIDS. Therefore, detection performance of the ML Ensemble NIDS via Forgo-The-Worst NIDS update rule for Case 4 is sensitive to the size of attack threshold choice, causing vulnerability and plausibility in deployment.

\begin{figure}[ht]
\vskip 0.2in
\begin{center}
 \centerline{\includegraphics[width=0.83\columnwidth]{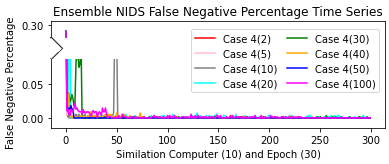}}
 \label{ieee-arnids-aide-s4}
 \caption{ML Ensemble NIDS Trend charts of FN ratio Series at each epoch and computer for Case 4.}
\end{center}
\vskip -0.2in
\end{figure}

Case 5 extends Case 4 but changes its NIDS update rule. Upon triggering an NIDS retraining request, Case 5 allows retraining and replacing all the active NIDS correspondingly, namely, Update-All-NIDSs rule (UALL), thereby retraining and updating the NIDS model $RF$, $HGI$ and $HGA$ respectively at the same run and time. As expected, its detection performance greatly deteriorates at the first computer and first epoch, delivering False Negative percentage at 29\%, as depicted in Fig. 9.
However, after completing the first NIDS retraining request to update and replace the whole ML Ensemble NIDS, the deterioration stops on the second computer at the first simulation epoch, and then it attains perfect detection score throughout the rest of simulation. Therefore, the sustained perfect detection performances of the ML Ensemble NIDS for Case 5 suggests that the Update-ALL-NIDS update rule mitigates the ML Ensemble NIDS model’s sensitivity to the size of attack threshold choice.

\begin{figure}[ht]
\vskip 0.2in
\begin{center}
 \centerline{\includegraphics[width=0.84\columnwidth]{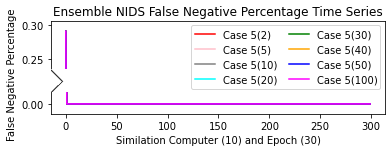}}
 \label{ieee-arnids-aide-s5}
 \caption{ML Ensemble NIDS Trend charts of FN ratio Series at each epoch and computer for Case 5.}
\end{center}
\vskip -0.2in
\end{figure}

Case 6 extends Case 5 to simulate production environment with no prior knowledge of the nature of network traffic. It uses hypergraph construct to detect if there exists any signature pattern of port scanning activities. Given no prior knowledge of the nature of network traffic, selecting the retrained NIDS with higher detection rate to update the active NIDS is in alignment of the commonly adopted zero-trust practice in cybersecurity. Its detection performance is the same as that of Case 5, namely, volatile initially but attaining 100\% detection level after updating the ML Ensemble NIDS at all the threshold choice in $TH$.

\subsection{Evaluating the Ensemble NIDS with CIC-IDS2017}

This section extends to evaluate the ML Ensemble NIDS with the second setting, namely the entire CIC-IDS2017 dataset and the set of adversarial examples generated, with Case 5 scenario that allows retraining and replacing all the active NIDS correspondingly. In addition to the Port Scan (260,293) and BENIGN (2,403,674) activities, there are thirteen (13) other types of attacks, comprising DoS Hulk (235,435), DoS GoldenEye (10,540), DoS slowloris (5,928), DoS Slowhttptest (5,635), DDoS (131,016), FTP-Patator (8,110), SSH-Patator (6,031), Bot (2,014), Web Attack - Brute Force (1,538), Web Attack - XSS (667), Web Attack - Sql Injection (22), Infiltration (36), and Heartbleed (11). \\



In this extension, the NIDS Model Module is accordingly modified to retain existing NIDS models if any of their detection performances in $F_1$ Score is higher than the best of the newly retrained NIDS models. In addition, the authors encoded the pairs of non-hacker IPs in each record of the CIC-IDS2017 dataset with a set of weights approximating the distribution of average $s$-closeness centrality (as depicted by the blue curve in Fig. 4) to help differentiate benign activities from other types of attacks. The set of weights approximating its $11$ features of average $s$-closeness centralities for any non-hacker pairs of IPs is:
\[ \left( 0.2, 0.15, 0.1, 0.05, 0.04, 0.03, 0.02, 0.01, 0.008, 0.006, 0.005 \right) \]

The total 2,668,152 benign and attack traffic (comprising of the 667,038 attacks and 2,001,114 BENIGN activities) were randomly sampled and split into 300 batches, roughly 8,900 records for each batch (comprising 25\% attacks and 75\% BENIGN activities) to feed into the simulation over a network of 10 users in 30 epochs. False negative percentage, accuracy, precision, recall and $F_1$ score were computed and evaluated throughout the simulation.

The numerical results summarized in Fig. 10 shows that the detection performance of the proposed ML Ensemble NIDS model based on $F_1$ score is not stable when the thresholds to trigger model re-training were set at $10$ attacks or lower. This suggests more frequent model retraining does not necessarily result in effective model learning. However, when the thresholds to trigger model re-training were set at $20$ attacks or higher, the results depicted Fig. 11 show that its detection performances are highly effective in $F_1$ Scores.


To illustrate the proposed ML Ensemble NIDS model's robustness against and resiliency to the attacks in the CIC-IDS2017 dataset, the baseline model results were obtained without using the hypergraph-based features in the Case 5 simulation setting that starts with a set of three identical NIDS models trained solely on NRF dataset and then follows with re-training NIDS NRF models separately. The baseline model results, as depicted in Fig. 12, improve initially but deteriorate eventually in most threshold choices in the simulation. Therefore, with proper choices of threshold, our numerical results provide evidence that the features of the $s$-closeness centralities of dynamic hypergraphs help the tree-based ML models, initially trained only on port scan activities, to learn and improve over time to give near perfect detection performances, furnishing both robustness and resiliency in the proposed ML Ensemble NIDS model.

\begin{figure}[ht]
\vskip 0.2in
\begin{center}
 \centerline{\includegraphics[width=0.8\columnwidth]{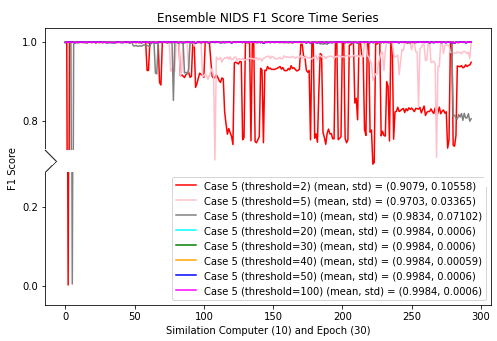}}
 \label{ieee-arnids-aide-s5-cicids-all-f1}
 \caption{ML Ensemble NIDS Trend charts of $F_1$ Score Series for each threshold choice in $TH$ at each epoch and computer for Case 5.}
\end{center}
\vskip -0.2in
\end{figure}



\begin{figure}[ht]
\vskip 0.2in
\begin{center}
\centerline{\includegraphics[width=0.85\columnwidth]{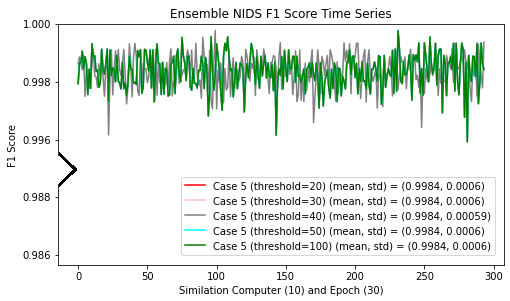}}
 \label{ieee-arnids-aide-s5-cicids-large-f1}
 \caption{ML Ensemble NIDS Trend charts of $F_1$ Score Series for the threshold choices  $\ge 20$ in $TH$ at each epoch and computer for Case 5.}
\end{center}
\vskip -0.2in
\end{figure}



\begin{figure}[ht]
\vskip 0.2in
\begin{center}
 \centerline{\includegraphics[width=0.8\columnwidth]{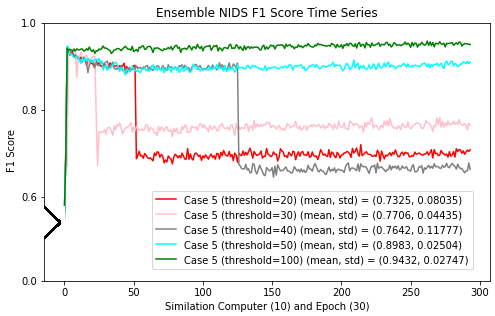}}
 \label{ieee-arnids-aide-s5-cicids-large-f1}
 \caption{ML Ensemble NIDS Trend charts of $F_1$ Score Series for the threshold choices  $\ge 20$ in $TH$ at each epoch and computer for Case 5.}
\end{center}
\vskip -0.2in
\end{figure}



\subsection{Numerical Results Summary}

Cyber-attacks are occurring across more heterogeneous and complex vectors. Numerical results indicate that the ML Ensemble NIDS had the best and near perfect performance over port scan dataset for Case 5 and 6 – comprising an ensemble of three NIDS along with Update-ALL-NIDS rule when needed. The significant improvement of the ML Ensemble NIDS indicates that the two sets of hypergraph-based features, namely $HGI$ and $HGA$, derived from the higher-level interactions that the hypergraph models are indeed meaningful. These settings also allow the ML Ensemble NIDS to be insensitive to the size of attack threshold choices for the reduced setting with port scan data only, as partially depicted in Fig. 13. The ML Ensemble NIDS' insensitivity to the threshold size retains at slightly higher values.

The consistent and near perfect $F_1$ Scores obtained from evaluating the ML Ensemble NIDS with the CIC-IDS2017 dataset illustrate that the uses of the derived hypergraph-based features of $s$-closeness centralities furnish the proposed model's robustness and resiliency. Specifically, the ML Ensemble NIDS stumbled at initial simulation epochs but bounced back once it learned the failures (i.e., exhibiting resiliency) and then maintained the high detection efficacy throughout the simulation (i.e., exhibiting robustness). As summarized in Table ~\ref{ensemble-nids-perf-eval-summ}, numerical results suggest that the ML Ensemble NIDS comprising $RF$ NIDS, $HGI$ NIDS and $HGA$ NIDS along with Update-ALL-NIDS update rule (namely, the settings for Case 5) exhibits robust potency in resolving the complex dynamics rendered by the combinations of adversarial examples, attack threshold choices, and production environment.

\begin{figure}[ht]
\vskip 0.2in
\begin{center}
 \centerline{\includegraphics[width=0.85\columnwidth]{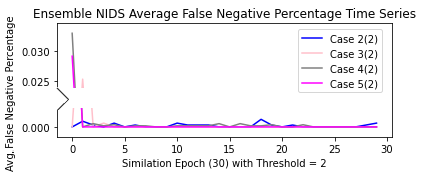}}
 \label{ieee-arnids-aide-s2-5}
 \caption{ML Ensemble NIDS Trend charts of average FN ratio Series for Case 2 through 5 with Threshold = 2.}
\end{center}
\vskip -0.2in
\end{figure}

\begin{table}[ht]
\caption{Performances Evaluation Summary of the ML Ensemble NIDS}
\label{ensemble-nids-perf-eval-summ}
\vskip 0.03in 
\begin{center}
\begin{small}
\begin{sc}
\begin{tabular}{lccc}
\hline
NIDS Update Rule & Adv. Ex. & Thresh. & Prod. \\
\hline
Static    & No & No & * \\
Forgo-The-Worst & No & No & * \\
Update-ALL-NIDS    & Yes & Yes & Yes \\
\hline
\end{tabular}
\\
\tiny{* no numeric results to report.}
\end{sc}
\end{small}
\end{center}
\vskip -0.1in
\end{table}

\section{Conclusions and Discussions}

This work advances the use of a hypergraph-based approach to devise a novel ML Ensemble NIDS detecting port scan attacks and adversarial examples with high efficacy. Specifically, the work derives a set of $s$-closeness centrality metrics capturing the adversary's signature pattern of port scan activities on a targeted machine. The work further investigates ways to incorporate these hypergraph metrics to render features augmenting the NIDS training dataset to train the effective proposed ML Ensemble NIDS.

Future research can extend these concepts from adding more hypergraph metrics to using such metrics to analyze other types of network intrusions. Non-tree based NIDS models may reduce the size of the attack threshold to trigger and allow for more frequent and effective learning in the model re-training. In light of differences in the evading performance level of adversarial examples between commonly adopted ML models and generated adversarial examples, several other approaches to generate adversarial examples based on evolutionary computation and deep learning will be investigated further to examine adversarial robustness of the ML Ensemble NIDS \cite{Alhajjar_Maxwell_Bastian_2021,Schneider_Aspinall_Bastian2021}. Adversarial examples can also be used to examine the effectiveness of the hypergraph-based ML Ensemble NIDS adversarial robustness. Other techniques such as incremental learning could also be explored to help build a self-sustaining, intelligent NIDS.



\bibliographystyle{IEEEtran}
\bibliography{arnids-aide-literature}


 





\end{document}